\documentclass[11pt]{article}

\usepackage[T1]{fontenc}
\usepackage{lmodern}
\usepackage{microtype}
\usepackage{geometry}
\usepackage{graphicx}
\usepackage{amsmath, amssymb}
\usepackage{booktabs}
\usepackage{tabularx}
\usepackage{array}
\usepackage{float}
\usepackage{caption}
\usepackage[percent]{overpic}
\usepackage[dvipsnames]{xcolor}
\usepackage{authblk}
\usepackage[
  backend=biber,
  style=phys,
  sorting=none,
  sortcites=true,
  natbib=true,
  date=year,
  doi=true,
  eprint=true,
  url=false,
  isbn=false
]{biblatex}
\AtEveryBibitem{%
  \clearfield{url}%
  \clearfield{urldate}%
  \ifentrytype{inproceedings}{\clearfield{eventtitle}}{}%
}
\renewbibmacro*{begentry}{\midsentence}
\DeclareBibliographyAlias{online}{article}
\DeclareFieldFormat[online]{title}{#1}
\usepackage[
  colorlinks=true,
  linkcolor=Magenta,
  citecolor=NavyBlue,
  urlcolor=Cyan
]{hyperref}

\title{\textbf{Continual-learning rules shape representational drift}}

\author[1,2]{Yikai Si}
\author[1,3]{Shanshan Qin\thanks{Corresponding author. \href{mailto:ssqin@sjtu.edu.cn}{ssqin@sjtu.edu.cn}}}
\affil[1]{School of Physics and Astronomy, Shanghai Jiao Tong University, Shanghai 200240, China}
\affil[2]{Zhiyuan College, Shanghai Jiao Tong University, Shanghai 200240, China}
\affil[3]{Institute of Natural Sciences, Shanghai Jiao Tong University, Shanghai 200240, China}

\date{}

\begin{document}

\maketitle


\begin{abstract}
Lifelong learning requires acquiring new knowledge without erasing the old. Yet neural population codes for familiar stimuli and behaviors change over days and weeks. This coexistence of stable memory and changing internal codes may reveal how a learning system prevents forgetting. We therefore tested whether different continual-learning mechanisms produce distinct patterns of representational drift. We trained convolutional neural networks on sequential image classification tasks and recurrent neural networks on sequences of cognitive tasks, tracking representations of fixed probe set across learning. Experience replay preserved earlier tasks in both architectures while representations drifted progressively with the number of intervening tasks. Drift was structured: later visual-processing stages and recurrent units' temporal tuning were especially labile, whereas coarse class organization and task-relevant temporal structure persisted. In contrast, algorithm that strongly anchored weights nearly froze representations. Directly anchoring an old representation during replay likewise suppressed drift and impaired acquisition of subsequent tasks. Together, these results link representational drift to the stability--plasticity trade-off: drift magnitude is shaped by the mechanism that protects old knowledge, and suppressing it can restrict future learning. Drift may therefore provide an observable signature of the constraints that enable continual learning in brains and machines.
\end{abstract}

\section{Introduction}

Brains retain memories and well-practiced behaviors for years, yet the neural activity supporting them is not fixed. Across brain regions and species, neuronal population activities that encode a familiar stimulus, environment, or action gradually change over days and weeks even when behavior remains stable. This phenomenon, termed representational drift, has been observed in many brain areas, such as the hippocampus \citep{ziv2013Longterm,geva2023Time,gonzalez2019Persistence,climer2025Hippocampal}, posterior parietal cortex \citep{driscoll2017Dynamic}, visual cortex \citep{deitch2021Representational,marks2021Stimulusdependent}, auditory cortex \citep{noda2025Homeostasis}, and piriform cortex \citep{schoonover2021Representational}. 
This apparent instability poses a fundamental puzzle: how can a downstream circuit produce a stable output from a code whose individual components continually change?

Population geometry provides part of the answer. Individual neurons can change their tuning while the population activity remains confined to a stable low-dimensional structure \citep{keinath2022Representation,qin2023Coordinated,sylte2025Coordinated,peters2026Coordinated}. Downstream circuits may therefore recover a consistent signal despite turnover among the neurons carrying it \citep{gallego2020Longterm,rule2022Selfhealing}. These observations have shifted the interpretation of drift: rather than simply reflecting biological noise or circuit degradation, drift may be a structured consequence of the processes that keep neural systems adaptive \citep{rule2019Causes,driscoll2022Representational}. What remains unclear is which learning processes generate such change while preserving memory and behavior.

Existing theories emphasize ongoing optimization, stochastic synaptic fluctuations, or a balance between the two. Continued training can slowly move a network among solutions with similar performance \citep{driscoll2022Representational,ratzon2024Representational}, whereas activity-independent synaptic volatility can continuously perturb tuning \citep{mongillo2017Intrinsic,qin2023Coordinated,rule2022Selfhealing,micou2026Statistics}. Recent works combine these mechanisms, treating drift as a balance between random synaptic change and learning-dependent restabilization \citep{eppler2026Representational,morales2025Representational}. Most such models, however, use a single-task setting: a network learns one task and is then updated or perturbed after performance has converged. Biological systems face a more demanding problem. They must preserve many old memories while acquiring new ones, precisely the regime in which representational drift and continual learning become inseparable.

This regime exposes the stability--plasticity dilemma. New learning requires internal change, but unconstrained change can overwrite existing knowledge. Drift might therefore be an incidental trace of weight updates \citep{vanderveldt2026Learning}, or it might help maintain flexibility, allocate resources to new memories, or counteract the loss of plasticity that develops over long training sequences \citep{ratzon2024Representational,dohare2024Loss}. Before asking whether drift is useful, however, one must determine which mechanisms of continual learning permit it and which suppress it. We therefore ask: does the learning rule leave a distinguishable signature in the amount and organization of representational drift?

Artificial neural networks provide a controlled testbed because their learning rules, weights, representations, and outputs are all observable. Previous works show that networks trained with noisy stochastic gradient descent could reproduce qualitative features of cortical drift, and the geometry of that drift depends on the perturbation used \citep{aitken2022Geometry,ratzon2024Representational}. By comparing neural code in networks trained with different continual learning algorithms, we can further reveal how different constraints on plasticity lead to different signature of drift. Parameter-regularization methods such as elastic weight consolidation (EWC) and synaptic intelligence protect knowledge by anchoring weights that are important for earlier tasks \citep{kirkpatrick2017Overcoming,zenke2017Continual}; these methods should suppress drift in the representations of old tasks. By contrast, experience replay interleaves stored examples from previous tasks while leaving the weights free to move \citep{chaudhry2019Tiny,vandeven2020Braininspired}. Functional-distillation and orthogonal-subspace methods constrain network outputs or activity subspaces rather than individual parameters \citep{li2018Learning,duncker2020Organizing}. These approaches should allow different degrees and geometries of drift. Although this prediction has been articulated in principle \citep{vanderveldt2026Learning}, it has not been tested directly across architectures and task domains.

Here, we test this in two systems. Convolutional neural networks (CNNs, ResNet-18) learned sequences of image-classification tasks drawn from ImageNet \citep{deng2009ImageNet}, and continuous-time recurrent neural networks (RNNs) learned a suite of cognitive tasks \citep{yang2019Task}. After each task we presented the same held-out probe set and measured how its representation had changed. In the CNNs we compared naive sequential training, EWC, Learning without Forgetting (LwF), and experience replay; in the RNNs, naive training and replay.

We found that replay preserved earlier tasks in both systems even as the task-1 representation drifted progressively with the number of intervening tasks. This drift was structured rather than diffuse. In CNNs, later processing stages changed faster than early ones while coarse class organization stayed recognizable; in the RNNs, the temporal tuning of individual units changed faster than their mean activity while the temporal organization of the trial persisted. What a learning rule constrained, and how strongly, set the scale of this change: LwF, which constrains the input--output function, produced intermediate drift, whereas strong EWC regularization nearly froze representations but also limited later learning. Anchoring the task-1 representation directly during replay reproduced that cost, suppressing drift and impairing acquisition of new tasks. The magnitude and organization of drift therefore report on the constraints that govern continual learning, linking how a system protects old knowledge to how much capacity it retains for new.

\section{Results}

We asked whether the rule used for continual learning leaves a characteristic signature in representational drift. We studied a CNN (ResNet-18) learning sequential image-classification tasks and a continuous-time RNN learning sequential cognitive tasks. For CNNs, we compared naive sequential training, EWC (parameter anchoring) \citep{kirkpatrick2017Overcoming}, LwF (functional distillation) \citep{li2018Learning}, and experience replay (rehearsal without direct parameter anchoring) \citep{chaudhry2019Tiny}. For the RNN, we compared naive training and replay method. After each task, we presented the same held-out probe set from task 1 and measured how its representation had changed. We evaluated three questions for each method: whether it retained earlier tasks, whether its representation drifted, and how drift was organized. 

\begin{figure}[H]
  \centering
  \begin{minipage}[c]{0.5\textwidth}
    \centering
    \begin{overpic}[width=0.95\linewidth]{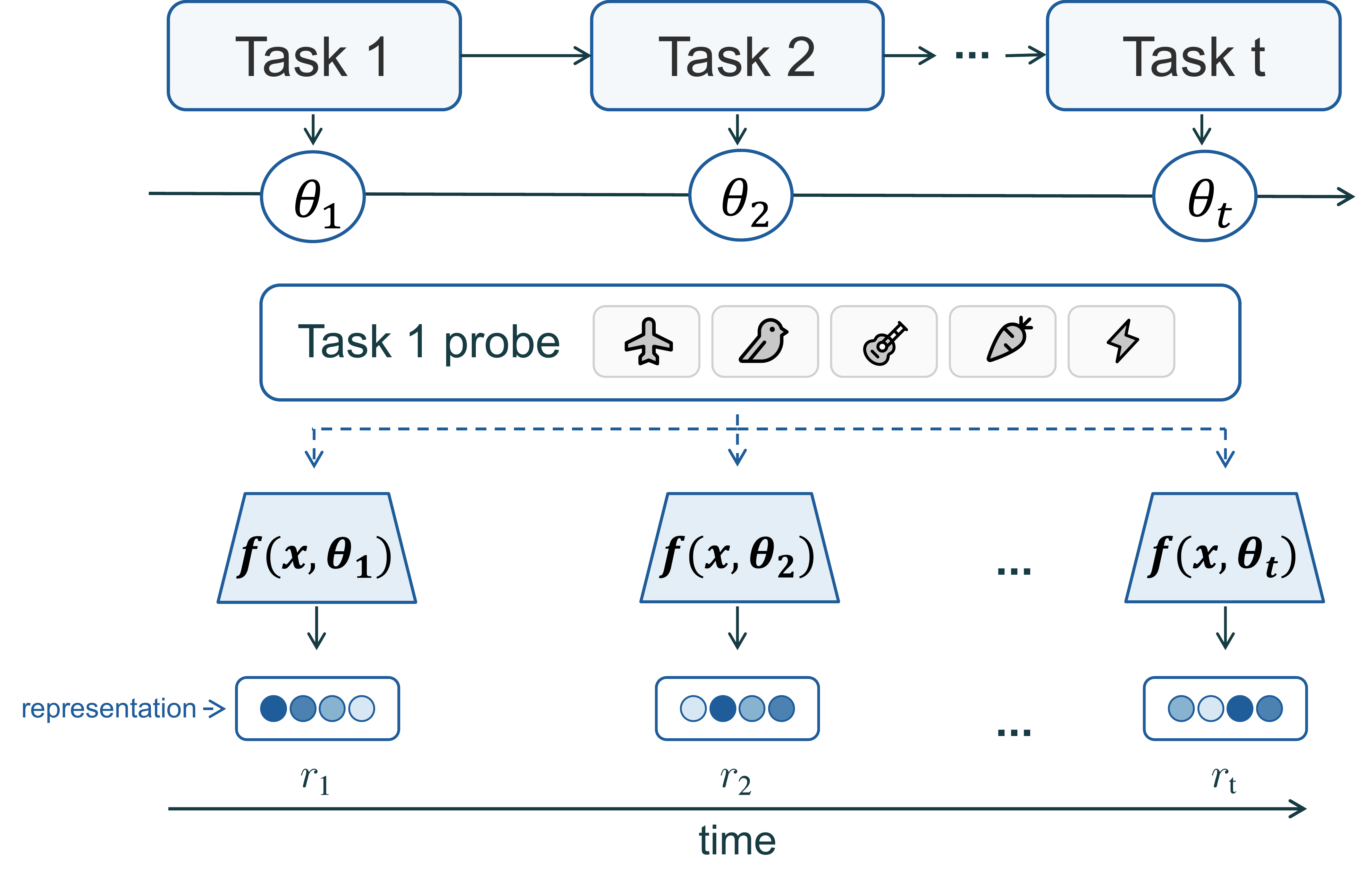}
      \put(1,61){{\textbf{\textsf{a}}}}
    \end{overpic}
  \end{minipage}%
  \begin{minipage}[c]{0.5\textwidth}
    \centering
    \begin{overpic}[width=0.95\linewidth]{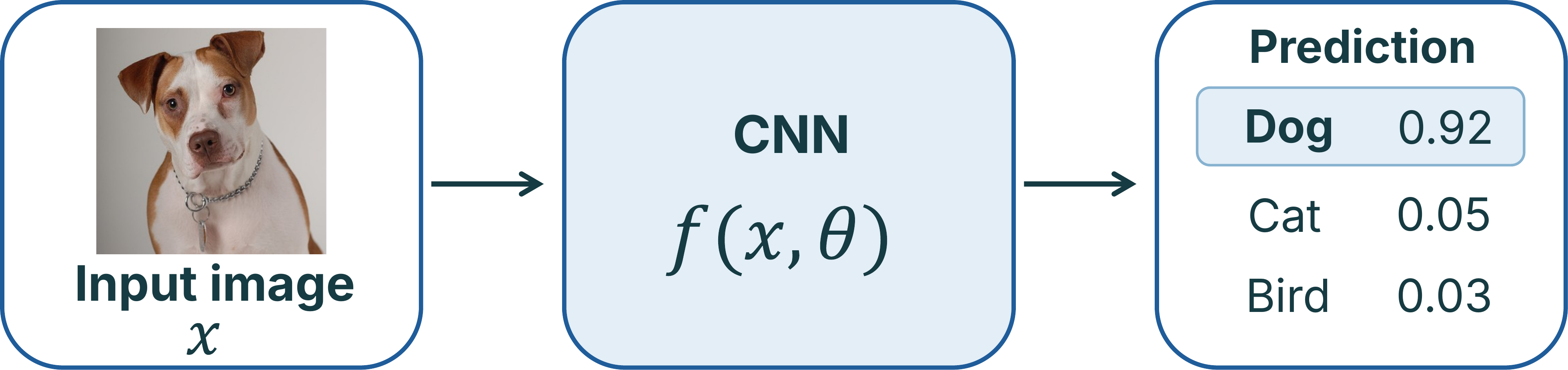}
      \put(-1,25){{\textbf{\textsf{b}}}}
    \end{overpic}\\[14pt]
    \begin{overpic}[width=0.95\linewidth]{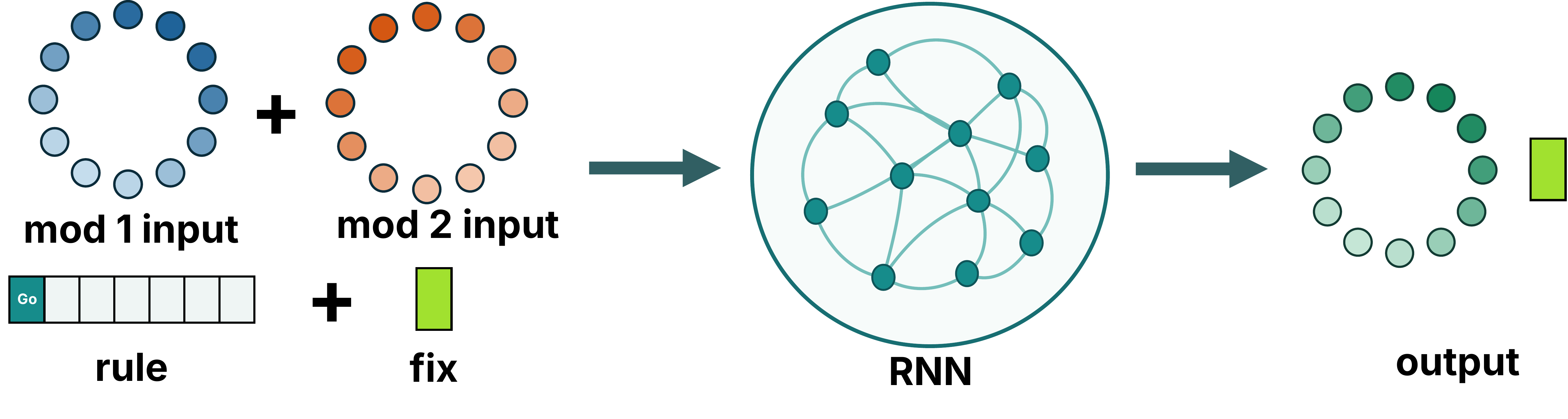}
      \put(-1,27){{\textbf{\textsf{c}}}}
    \end{overpic}
  \end{minipage}
  \caption{Overview of the experimental design. (a) Continual-learning and probe-drift schematic: the network is trained sequentially across tasks (Task~1, Task~2, \dots, Task~$t$), producing checkpoint weights $\theta_1,\theta_2,\dots,\theta_t$; a fixed probe set drawn from Task~1 is passed through the network at every checkpoint to obtain the representations $r_1,r_2,\dots,r_t$ used to track representational drift over time. (b) Image-classification task: a convolutional network (ResNet-18) maps an input image to a category prediction. (c) Cognitive-task pipeline: a continuous-time recurrent network maps a modality input, task-rule cue, and fixation signal to a behavioral output.}
  \label{fig:overview}
\end{figure}

\subsection{Drift in hierarchical visual representations}

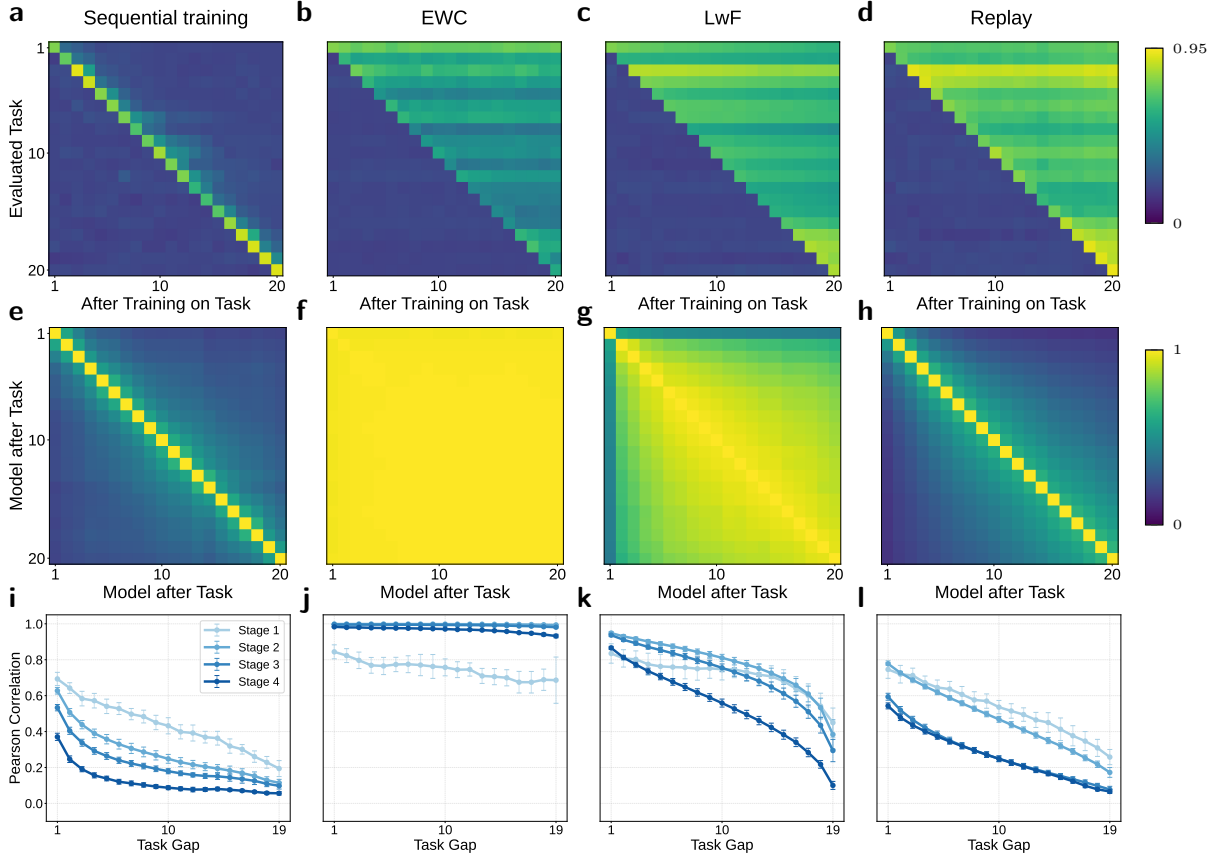
\begin{figure}[ht]
  \centering
  \begin{overpic}[width=\linewidth]{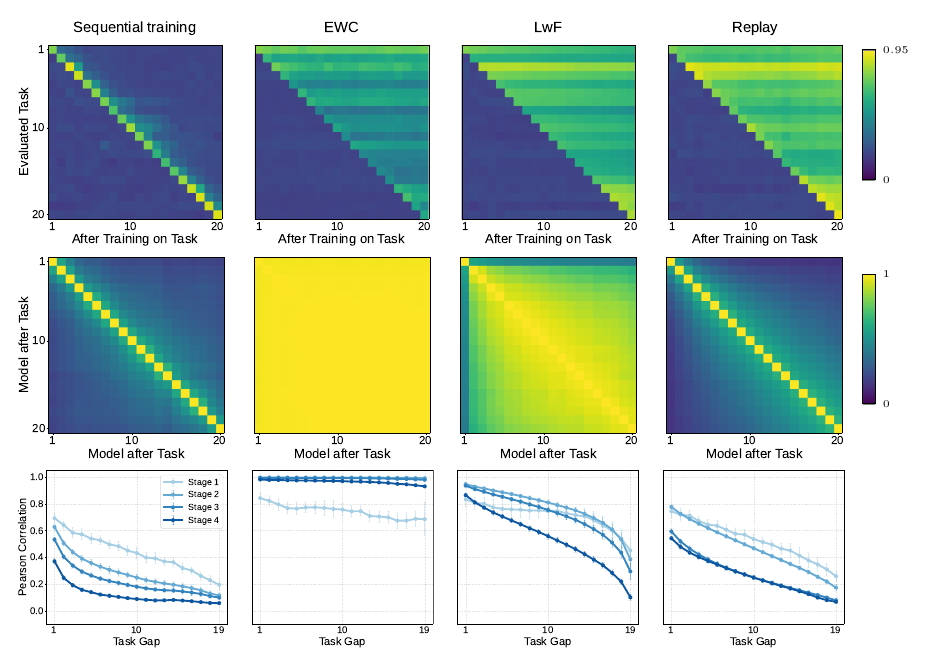}
    \put(2,68){{\textbf{\textsf{a}}}}
    \put(25,68){{\textbf{\textsf{b}}}}
    \put(47.5,68){{\textbf{\textsf{c}}}}
    \put(70,68){{\textbf{\textsf{d}}}}
    \put(2,44){{\textbf{\textsf{e}}}}
    \put(25,44){{\textbf{\textsf{f}}}}
    \put(47.5,44){{\textbf{\textsf{g}}}}
    \put(70,44){{\textbf{\textsf{h}}}}
    \put(2,21){{\textbf{\textsf{i}}}}
    \put(25.5,21){{\textbf{\textsf{j}}}}
    \put(47.5,21){{\textbf{\textsf{k}}}}
    \put(70,21){{\textbf{\textsf{l}}}}
  \end{overpic}
  \caption{ImageNet CNN summary across continual-learning methods. Columns correspond to methods (naive sequential training, EWC, LwF, Replay); rows show task accuracy, pairwise model similarity at residual stage 3, and sample-PV Pearson correlation as a function of task gap. Panels: (a,e,i) naive sequential training, (b,f,j) EWC, (c,g,k) LwF, and (d,h,l) Replay.}
  \label{fig:cnn_method_grid}
\end{figure}

The four methods used in CNNs occupied distinct positions in the trade-off among retention, plasticity, and representational stability (Fig.~\ref{fig:cnn_method_grid}). Naive sequential training learned each new task but rapidly lost earlier ones, producing high accuracy mainly along the diagonal of the task-by-checkpoint accuracy matrix (Fig.~\ref{fig:cnn_method_grid}a). EWC showed the opposite pattern: it kept checkpoint representations nearly unchanged, but the network adapted poorly to later tasks (Fig.~\ref{fig:cnn_method_grid}b, f, and j). LwF was intermediate, retaining earlier tasks better than naive training while continuing to learn new ones (Fig.~\ref{fig:cnn_method_grid}c, g, and k). Replay provided the strongest overall retention, maintaining high performance on earlier tasks while learning across the sequence (Fig.~\ref{fig:cnn_method_grid}d, h, and l).

Yet performance retention did not require the probe representation to remain fixed. Under replay, the network learned and retained the task sequence, yet pairwise checkpoint similarity declined smoothly with task separation and sample-level population-vector correlations decreased in every probed stage (Fig.~\ref{fig:cnn_method_grid}d, h, and l). Because task-1 performance remained high, this change met the defining condition for representational drift: a changing internal code accompanied by stable behavior. LwF also produced gradual drift, at an intermediate rate. EWC instead approached a freezing limit, but its representational stability coincided with poor acquisition of later tasks (Fig.~\ref{fig:cnn_method_grid}b, f, and j). We therefore treated EWC as a strongly constrained reference rather than a performance-matched comparison. Among the methods that both learned and showed representational drift, later layers changed more rapidly than earlier layers, consistent with greater lability of the more task-specific visual features.

\begin{figure}[ht]
  \centering
  \begin{overpic}[width=0.92\textwidth]{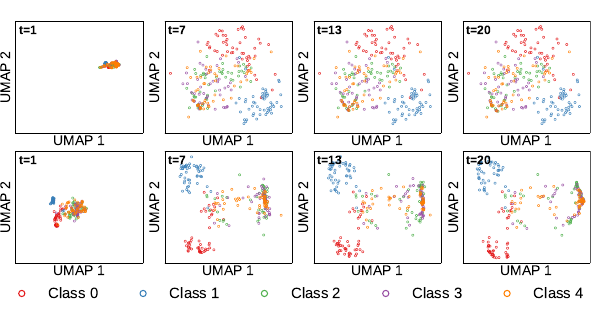}
    \put(-1,49){{\textbf{\textsf{a}}}}
    \put(-1,27){{\textbf{\textsf{b}}}}
  \end{overpic}
  \caption{ImageNet probe representations at layer 4. Activations from all checkpoints
  were compressed by principal component analysis (PCA) and embedded in a shared uniform
  manifold approximation and projection (UMAP) space. Points are colored by task-1 class;
  columns show successive checkpoints. Rows show (a) LwF and (b) replay.}
  \label{fig:cnn_umap}
\end{figure}

To visualize what changed and what persisted, we projected probe activations of layer-4 from all checkpoints onto a shared two-dimensional plane using principal component analysis (PCA) followed by uniform manifold approximation and projection (UMAP; Methods). Under replay, task-1 classes remained visibly clustered at each checkpoint even as the locations and shapes of the clusters changed; LwF showed an intermediate degree of cluster preservation (Fig.~\ref{fig:cnn_umap}). These embeddings illustrate how coarse class organization can coexist with movement in feature space. Note that UMAP is nonlinear, and can only provide a qualitative view. A quantitative comparison of sample-wise similarity matrices by centered kernel alignment shows that this relational geometry is much more stable than the individual population vectors (Appendix, Fig.~\ref{fig:cka_similarity_gap}).

\begin{figure}[ht]
  \centering
  \begin{minipage}[t]{0.47\textwidth}
    \centering
    \begin{overpic}[width=\linewidth]{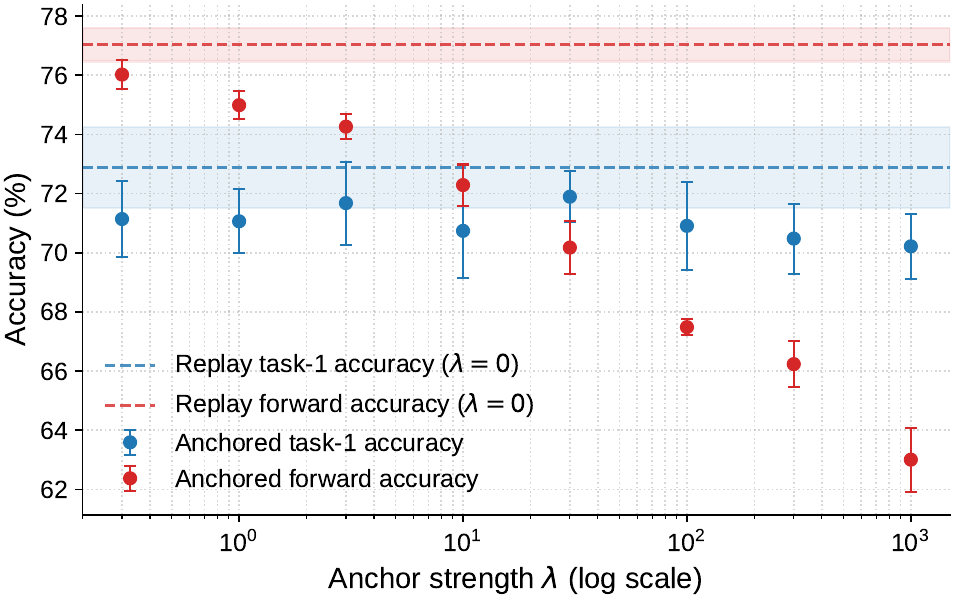}
      \put(-2,64){{\textbf{\textsf{a}}}}
    \end{overpic}
  \end{minipage}
  \hfill
  \begin{minipage}[t]{0.47\textwidth}
    \centering
    \begin{overpic}[width=\linewidth]{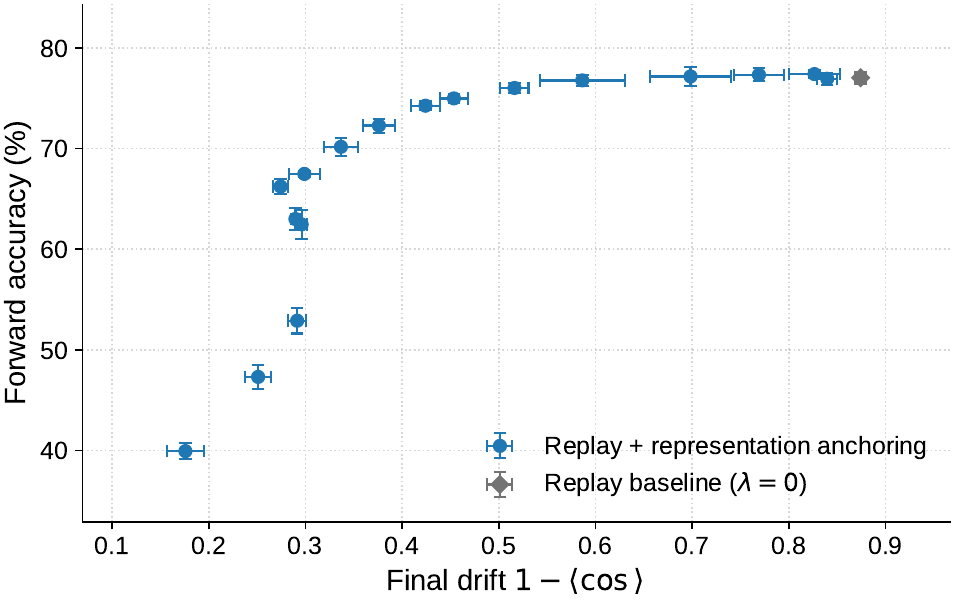}
      \put(-2,64){{\textbf{\textsf{b}}}}
    \end{overpic}
  \end{minipage}
  \caption{Trade-off between representational stability and plasticity in replay method
  .
  (a) Task-1 accuracy and forward accuracy as a function of representation-anchor strength
  $\lambda$. Dashed lines and shaded bands show the mean and 95\% confidence interval for
  unanchored replay ($\lambda=0$). (b) Forward accuracy versus final representational drift,
  measured as $1-\langle\cos\rangle$, across anchor strengths; the unanchored replay
  baseline is shown separately.}
  \label{fig:anchoring_tradeoff}
\end{figure}

The above results suggest that drift is correlated with the network's capacity to continue learning new tasks. To test this directly, we manipulated representational stability within replay by adding a penalty that anchored the task-1 representation (Methods). Increasing the anchor strength progressively reduced final drift while preserving task-1 accuracy, but it also reduced mean accuracy on newly introduced tasks, i.e., forward learning accuracy (Fig.~\ref{fig:anchoring_tradeoff}a). Across anchor strengths, larger final drift was associated with higher forward accuracy (Fig.~\ref{fig:anchoring_tradeoff}b). Thus, within a single replay-based regime, strongly restricting movement of an old representation impaired the network's capacity to acquire later tasks. This result reinforces the idea that representational drift is compatible with stable behavior and continual learning, while strong constraints on the representation can hinder plasticity and learning of new tasks. Although the intervention does not establish that drift is universally beneficial, it directly demonstrates a trade-off between representational stability and continued plasticity under the conditions tested.

\subsection{Drift in recurrent cognitive dynamics}
\begin{figure}[ht]
  \centering
  \begin{overpic}[width=\linewidth]{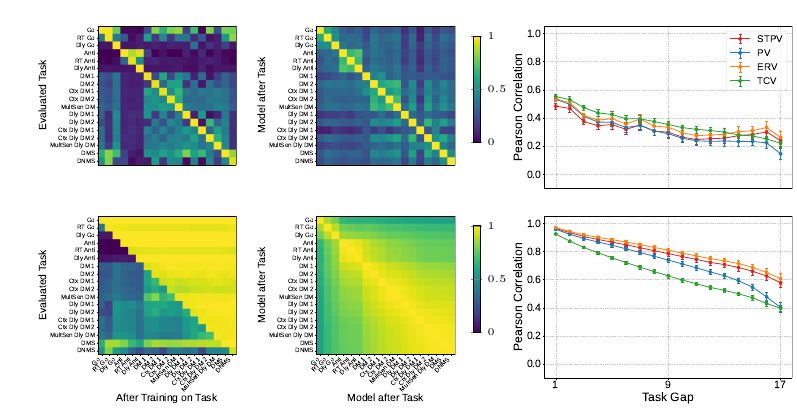}
    \put(3,49){{\textbf{\textsf{a}}}}
    \put(31,49){{\textbf{\textsf{b}}}}
    \put(63,49){{\textbf{\textsf{c}}}}
    \put(3,25){{\textbf{\textsf{d}}}}
    \put(31,25){{\textbf{\textsf{e}}}}
    \put(63,25){{\textbf{\textsf{f}}}}
  \end{overpic}
  \caption{RNN summary across continual-learning methods. Rows correspond to methods (naive sequential training, Replay); columns show task accuracy, pairwise STPV Pearson similarity for the fdgo probe task, and STPV, PV, ERV, and TCV correlations as a function of task gap. Panels: (a--c) naive sequential training and (d--f) Replay; within each row, columns show accuracy, similarity, and vector drift.}
  \label{fig:rnn_method_grid}
\end{figure}
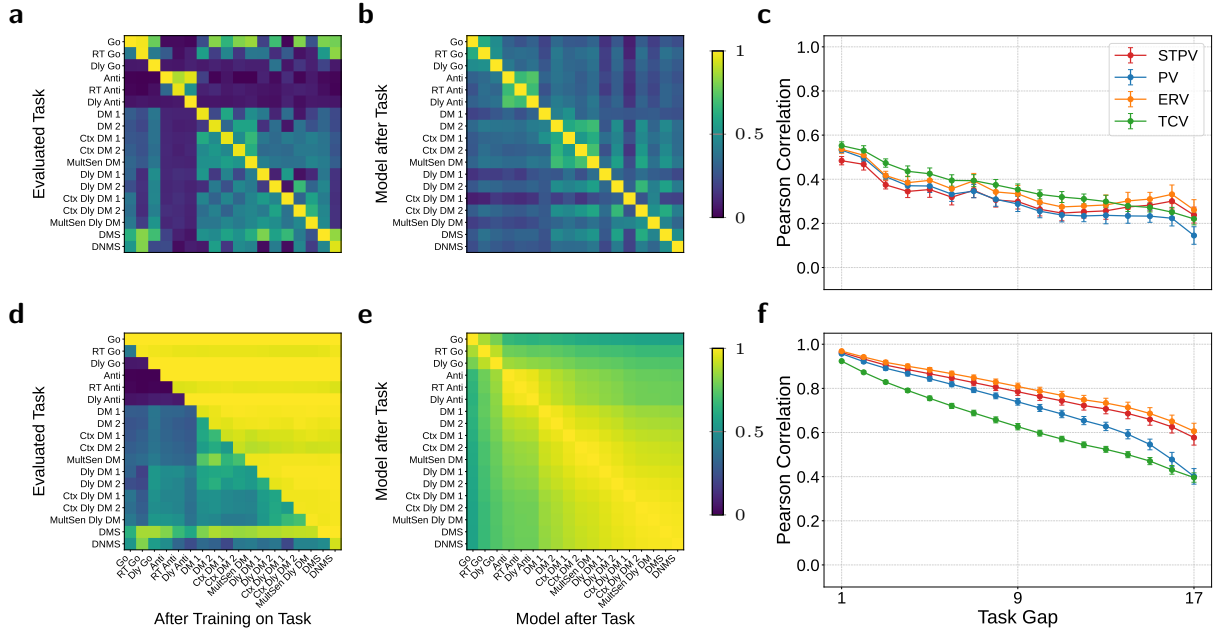

The above analysis concerned static, layerwise activation patterns in CNNs. We next asked whether stable behavior could coexist with drift in a dynamic representation: a temporally extended trajectory through recurrent state space. We trained a RNN sequentially on 18 cognitive tasks and compared naive training with replay (Fig.~\ref{fig:rnn_method_grid}).

Naive sequential training produced catastrophic interference: most off-diagonal entries in the task-accuracy matrix were near chance, indicating that each new task largely overwrote solutions for earlier tasks (Fig.~\ref{fig:rnn_method_grid}a). Replay instead produced near-perfect retention, with almost every task-by-checkpoint accuracy at least 0.95 (Fig.~\ref{fig:rnn_method_grid}d). The few exceptions reflected transient interference between task families, such as Go tasks tested immediately after introduction of Anti tasks.

This stable performance again coexisted with ongoing representational change. From each probe trial's hidden-state trajectory, we extracted four complementary summaries (Methods): the population vector (PV; all units at one time point), spatiotemporal population vector (STPV; the full trial trajectory), ensemble rate vector (ERV; each unit's trial-averaged activity), and tuning curve vector (TCV; each unit's time course). Under replay, STPV similarity between checkpoints decreased monotonically with task gap, reaching 0.59 at the maximum separation (Fig.~\ref{fig:rnn_method_grid}e). All four measures drifted, but at different rates (Fig.~\ref{fig:rnn_method_grid}f): the TCV was most labile, falling to approximately 0.43 at gap 17, whereas the ERV was most stable at approximately 0.63. The ordering ERV $>$ STPV $>$ PV $>$ TCV indicates that mean activity was better preserved than the precise temporal tuning of individual units---a dynamic analogue of the layer-dependent drift observed in the convolutional network.

Finally, we localized drift within the trial by comparing population states across both checkpoints and time points, separately for fixation and for the stimulus--response epochs. Under naive training, states from distant checkpoints were nearly uncorrelated, consistent with catastrophic overwriting. Under replay, states belonging to the same trial epoch remained substantially more similar, although their similarity declined gradually with checkpoint separation. The recurrent code therefore changed without losing the temporal organization needed for the task (Appendix, Fig.~\ref{fig:rnn_temporal_drift}).

\section{Discussion}

Our results connect representational drift to the central challenge of continual learning: preserving existing knowledge while remaining able to learn. Across feedforward and recurrent networks, experience replay maintained performance on earlier tasks even as their internal representations changed progressively. The form of this change was structured. Later visual-processing stages drifted more than earlier stages, the temporal tuning of recurrent units drifted more than their mean activity, and coarse class and trial-epoch organization remained recognizable. Continual-learning rules also produced distinct drift profiles: functional distillation allowed intermediate drift, whereas strong weight anchoring approached representational freezing but limited later learning. Finally, suppressing drift directly within replay impaired acquisition of subsequent tasks. Together, these findings show that representational drift is closely coupled to the constraints a learning system places on plasticity.

\paragraph{Stable behavior can coexist with changing internal codes.} Replay provides the clearest example of this dissociation. In both architectures, the networks retained earlier tasks while similarity between their representations declined with the number of intervening tasks. The recurrent-network results extend this conclusion from static feature vectors to temporally extended trajectories. Mean unit activity was relatively stable, whereas the temporal tuning of individual units was more labile; at the population level, the temporal organization of the task remained recognizable. This combination of stable output, changing individual components, and partially preserved population structure parallels the central empirical features of biological drift \citep{driscoll2022Representational}. It also exposes a limitation of evaluating continual-learning algorithms by accuracy alone: methods with comparable retention can differ substantially in how they reorganize their internal codes.

\paragraph{Learning rules leave distinct drift profiles.} The continuous learning methods used in CNN occupied different positions along axes of retention, plasticity, and representational stability. LwF constrained the network's input--output function while allowing its parameters and features to change. EWC directly constrained important parameters and, at the strong penalty used here, approached a freezing limit. Replay constrained performance on stored examples without anchoring a particular weight configuration. Their resulting drift profiles support the general prediction that both the target and strength of a learning constraint shape internal representational change. This comparison is qualitative rather than a ranking of algorithms, because the methods were not matched for old-task retention, new-task learning, or total parameter movement. Such matched comparisons will be needed to identify which property of an algorithm best predicts its drift profile.

\paragraph{Representational stability trades off with continued plasticity.} To vary representational stability without changing the learning rule, we introduced a representation-anchor penalty within replay. Stronger anchoring reduced movement of the task-1 representation while leaving task-1 performance stable, but it also reduced accuracy on newly introduced tasks. Forcing an old representation unchanged therefore restricts the network's ability to acquire later tasks: representational drift cannot always be removed from a continually learning system without affecting plasticity.

\paragraph{Implications for theories of representational drift.} Most mechanistic accounts study a circuit trained on a single task and then driven beyond convergence by continued optimization, synaptic noise, or both \citep{ratzon2024Representational,mongillo2017Intrinsic,qin2023Coordinated,eppler2026Representational,morales2025Representational}. Our experiments extend the continued-learning account to a multi-task regime in which new knowledge is acquired while old knowledge is actively maintained. They show that gradual, structured drift can emerge from ordinary task-driven learning without explicitly imposed synaptic fluctuations, and that replay can prevent this change from becoming catastrophic forgetting. The simulations do not identify the microscopic source of biological drift: different drift magnitudes could arise from the form of the learning constraint, the distance traveled in parameter space, stochastic fluctuations, or interactions among these factors. They do, however, establish continual-learning rules as a plausible and testable source of variation in drift.

\paragraph{Implication for biological learning.} Replay generated several qualitative features associated with cortical drift: change accumulated with intervening learning, drift rates differed across processing stages or representational components, and coarse task structure persisted despite turnover \citep{sylte2025Coordinated,noda2025Homeostasis}. This is consistent with theories in which repeated reactivation supports memory consolidation \citep{mcclelland1995Why}. The inference is one of algorithmic compatibility, not biological mechanism. Artificial experience replay is not a model of the cellular processes underlying hippocampal replay, and the results neither show that biological replay causes cortical drift nor exclude other plasticity rules. A stronger biological test would compare the geometry and rate of model drift quantitatively with longitudinal neural recordings and ask whether perturbing replay or plasticity changes those signatures as predicted.

\paragraph{Origin versus function.} Our primary question concerns which learning regimes permit drift, not whether drift has an adaptive function. The anchoring experiment strengthens the connection between representational flexibility and subsequent learning, but the penalty itself changes the optimization problem and may restrict useful degrees of freedom for reasons not specific to biological drift. Nor does the result exclude an anchoring scheme that could suppress drift at a smaller cost. The observed relationship is therefore consistent both with drift supporting continued plasticity and with drift being an inseparable consequence of the parameter changes required for new learning. Distinguishing these possibilities will require interventions that vary drift while matching retention, new-task performance, and parameter displacement. Such experiments could test whether drift actively counters the loss of plasticity that emerges during long task sequences \citep{dohare2024Loss}.

\paragraph{Limitations.} First, we used a task-incremental protocol in which task identity was available at evaluation; class-incremental learning may produce different drift profiles \citep{vandeven2022Three}. Second, the strongly regularized EWC regime learned later tasks poorly, so it is a freezing reference rather than a performance-matched control. Third, EWC and LwF did not achieve sufficiently reliable continual-learning performance in the RNN for an interpretable drift comparison; whether the CNN ordering generalizes to recurrent systems remains unknown. Finally, the reported profiles were averaged over independently trained networks per condition but without varying task order. Comparisons with biological drift must also account for differences in task structure, timescale, and recording statistics.

\paragraph{Outlook.} Representational drift provides a unique dimension for comparing continual-learning systems that behavioral accuracy alone cannot reveal. Systematic comparisons among parameter-, function-, and activity-constraining methods---matched for performance retention, plasticity, and parameter movement---could determine which algorithmic constraints govern the rate and geometry of drift. Combined with targeted interventions and quantitative comparison to longitudinal neural recordings, this approach could establish when drift is merely tolerated, when it enables future learning, and when it reveals the mechanisms by which a system balances memory with lifelong plasticity.

\section{Materials and Methods}

The study comprised two experimental tracks: a convolutional neural network (CNN) trained on sequential image-classification tasks and a recurrent neural network (RNN) trained on sequential cognitive tasks. We first describe the CNN (Sec.~\ref{sec:cnn_setup}) and RNN (Sec.~\ref{sec:rnn_setup}) experiments, then define the geometric measures used to quantify representational drift in both settings (Sec.~\ref{sec:notation}).

\subsection{CNN experimental setup}
\label{sec:cnn_setup}

\paragraph{Dataset and task structure.} The CNN experiment uses the first 100 classes of a processed ImageNet-1k dataset \citep{deng2009ImageNet,russakovsky2015ImageNet}, divided into $T=20$ sequential tasks of 5 classes each. Images are resized to $224\times224$ pixels. Training uses randomly resized crops with a scale range of 0.7--1.0 and horizontal flips. Validation and test images are resized and center-cropped. All images are normalized using ImageNet channel statistics.

\paragraph{Model architecture.} The CNN experiments use ResNet-18 backbones trained from scratch \citep{he2016Deep}. We replace Batch Normalization (BN) with Group Normalization (GN) \citep{wu2018Group} and Weight Standardization (WS) \citep{qiao2019MicroBatch}. Unlike BN, which accumulates running means and variances from the sequence of training batches, GN computes its normalization statistics separately for each image over groups of channels. This makes the normalization independent of the order in which task distributions are encountered and prevents shifts in stored BN statistics from confounding measurements of representational drift.

For a convolutional weight tensor $W\in\mathbb{R}^{C_{\mathrm{out}}\times C_{\mathrm{in}}\times k\times k}$, where $C_{\mathrm{out}}$ and $C_{\mathrm{in}}$ are the numbers of output and input channels and $k$ is the spatial kernel size, WS normalizes each output-channel filter before convolution. Let $N=C_{\mathrm{in}}k^2$ and let $c=1,\ldots,C_{\mathrm{out}}$ index the output channel and $j$ index the $N$ elements of a filter. The standardized weights are
\begin{equation}
  \mu_c=\frac{1}{N}\sum_{j=1}^{N}W_{c,j},
  \qquad
  \sigma_c=\sqrt{\frac{1}{N}\sum_{j=1}^{N}(W_{c,j}-\mu_c)^2},
  \qquad
  \hat W_{c,j}=\frac{W_{c,j}-\mu_c}{\sigma_c+\epsilon},
\end{equation}
where $\epsilon = 10^{-5}$ is used for numerical stability.

\textbf{ResNet-18-GN/WS.} It has a standard ImageNet-style stem: a $7\times7$ convolution with stride 2 followed by max-pooling. The four residual stages have 64, 128, 256, and 512 channels. We refer to
their outputs as stages 1--4, ordered from shallow to deep.  All convolutional layers use WS, and all normalization layers use GN. The number of groups is the largest divisor of the channel count that does not exceed 32.  After normalizing activations
within each group, GroupNorm applies a learnable affine transform
$\gamma\odot\hat{\mathbf x}+\boldsymbol\beta$, where $\hat{\mathbf x}$ denotes the
normalized activations, $\odot$ is element-wise multiplication, and
$\gamma,\boldsymbol\beta\in\mathbb{R}^{C}$ are per-channel scale and shift parameters. The scale parameter $\gamma$ of the second GN layer in each residual block is initialized to zero, so each block initially approximates an identity mapping. Convolutional weights use Kaiming normal initialization \citep{he2015Delving}, and the classifier is a single linear layer. The full network is trained end-to-end on every task.

\paragraph{Continual-learning methods.} All CNN experiments use a task-incremental protocol. During training and evaluation, the task identity is known, and the relevant output logits correspond to five classes. This protocol isolates changes in internal representations from the output ambiguity of class-incremental learning. Replay batches contain current and stored examples, so their training loss includes all classes seen up to the current task. Evaluation remains task-incremental.

We compare four methods. \emph{Naive sequential training} optimizes only the current-task loss and provides no explicit protection against forgetting. 

In \emph{Elastic weight consolidation }(EWC) \citep{kirkpatrick2017Overcoming}, the diagonal of Fisher information metric, $F_q$, is estimated from the training data after task $q$. It measures the importance of each parameter to that task. When learning task $t$, EWC minimizes
  \begin{equation}
    \mathcal{L}_{\mathrm{EWC}}(\theta) = \mathcal{L}_{\mathrm{CE}}(\theta) +
    \frac{\lambda}{2}\sum_{q<t}\sum_k F_q^{(k)}
    \bigl(\theta_k - \theta_{q,k}^*\bigr)^2,
    \quad \lambda > 0.
  \end{equation}
Here, $\theta_{q,k}^*$ is parameter $k$ after task $q$. The ImageNet experiment uses $\lambda=5\times10^5$ and retains a separate penalty for every previous task.

\emph{Experience replay} \citep{chaudhry2019Tiny} stores a uniformly sampled subset of previous training examples in a memory buffer $\mathcal{M}$. Training interleaves current-task batches with batches sampled from $\mathcal{M}$. The ImageNet and Tiny ImageNet buffers contain 300 examples per class; the CIFAR-100 buffer contains 400 examples per class.

To test whether suppressing representational drift impairs later learning,
we also study \emph{replay with representation anchoring}. After task~1, we cache
activations $\mathbf{r}_1(x)$ of a fixed probe set $\mathcal{P}$ drawn from that task
at residual stages 3 and 4. On subsequent tasks the training objective is
\begin{equation}
  \mathcal{L}_{\mathrm{anchor}}(\theta)
  = \mathcal{L}_{\mathrm{replay}}(\theta)
  + \lambda_{\mathrm{a}}\,
    \Bigl\langle
      d\!\bigl(\mathbf{r}_\theta^\ell(x),\mathbf{r}_1^\ell(x)\bigr)
    \Bigr\rangle_{x\in\mathcal{P},\,\ell},
\end{equation}
where the average runs over probe samples and stages $\ell\in\{3,4\}$, and
$d$ is a per-stage mean-squared displacement normalized by the mean squared norm of
$\mathbf{r}_1^\ell$. Unlike EWC, the penalty constrains representations rather than
parameters. We vary the anchor strength $\lambda_{\mathrm{a}}$ to examine the trade-off
between drift suppression and plasticity of learning later tasks.

In \emph{Learning without Forgetting} (LwF) \citep{li2018Learning}, the model is copied before task $t$ is learned. The copied model produces fixed logits $z^{\mathrm{old}}$ on the new-task data, and the updated model produces $z^{\mathrm{new}}$.We adapt LwF to our single-head task-incremental
setting: the new-task loss is cross-entropy on the five classes of the current task,
and distillation is applied only to previously seen classes. Let
$C_{t-1}=5(t-1)$ be the number of classes observed before task~$t$. Training minimizes
  \begin{equation}
    \mathcal{L}_{\mathrm{LwF}}(\theta)
    = \mathcal{L}_{\mathrm{new}}(\theta)
    + \lambda_{\mathrm d}\,
    \kappa_{\mathrm d}^{2}\,
    \mathrm{KL}\!\left[
      \operatorname{softmax}\!\left(\frac{{z^{\mathrm{old}}_{1:C_{t-1}}}}{\kappa_{\mathrm d}}\right)
      \,\middle\|\,
      \operatorname{softmax}\!\left(\frac{z^{\mathrm{new}}_{1:C_{t-1}}}{\kappa_{\mathrm d}}\right)
    \right],
  \end{equation}
where $\mathcal{L}_{\mathrm{new}}$ denotes the current-task
cross-entropy, $z_{1:C_{t-1}}$ is the corresponding logit slice,
$\lambda_{\mathrm d}=30$ is the distillation weight, and
$\kappa_{\mathrm d}=2$ is the temperature. The factor
$\kappa_{\mathrm d}^{2}$ follows the standard knowledge-distillation scaling of the
softened Kullback--Leibler divergence (KL) term \citep{hinton2015Distilling}.

\paragraph{Training.} All CNN experiments use stochastic gradient descent with Nesterov momentum of 0.9 and cosine learning-rate annealing within each task. Weight decay of $5\times10^{-4}$ is applied only to weight tensors. The optimizer state and learning rate are reset at the start of each task. At the end of each epoch, we evaluate the validation
loss. Training stops if this loss fails to decrease for 10 consecutive epochs;
we then restore the weights from the epoch with the lowest validation loss. Early stopping monitors validation loss with a patience of 10 epochs. ImageNet is trained for at most 60 epochs per task with an initial learning rate of 0.2 and a batch size of 256. A model checkpoint is saved after every task. To quantify a network's plasticity in the course of learning, we measure \emph{forward accuracy},  defined as the mean best validation accuracy attained on each newly introduced task while that task is
being trained. Task-1 accuracy denotes accuracy on the first task evaluated at the final
checkpoint.

\subsection{RNN experimental setup}
\label{sec:rnn_setup}

\paragraph{Cognitive tasks.} Yang et al.\ \citep{yang2019Task} define 20 trial-based cognitive tasks. We use 18 of them, excluding DMC-Go and DMC-NoGo. Those two tasks draw stimulus directions from a small discrete set, creating substantial overlap between independently generated training and test trials and making generalization difficult to estimate. The remaining tasks belong to four families: Go/Anti (6 tasks), Decision Making (5 tasks), delayed Decision Making (5 tasks), and Matching (2 tasks). Each trial follows the sequence fixation $\to$ stimulus $\to$ optional delay $\to$ response.

\paragraph{Trial generation and network interface.} Trials are generated procedurally as detailed in \citet{yang2019Task}. A batch contains $S$ time steps of duration $\Delta t=20\,\mathrm{ms}$; $S$ may vary across batches because the durations of the fixation, stimulus, delay, and response epochs are sampled from task-specific ranges. At each time step, the input $\mathbf{x}_s\in\mathbb{R}^{83}$ contains one fixation unit, two sensory rings of 32 units each, and 18 rule units. The fixation unit is 1 during fixation and 0 during the response epoch. The rule units form a one-hot code for the current task.

Each sensory ring represents a direction $\theta\in[0,2\pi)$ with 32 uniformly spaced preferred directions $\theta_i$. Let
\begin{equation}
  d_{\mathrm{circ}}(\theta,\theta_i)
  = \min\!\left(|\theta-\theta_i|,\,2\pi-|\theta-\theta_i|\right)
\end{equation}
be the circular angular distance. The activity of sensory unit $i$ is
\begin{equation}
  x_i(\theta)=0.8\exp\!\left[
    -\frac{1}{2}\left(\frac{8d_{\mathrm{circ}}(\theta,\theta_i)}{\pi}\right)^2
  \right].
\end{equation}
Stimulus strength and coherence are sampled from task-specific distributions. Independent Gaussian input noise with standard deviation $\sigma_x\sqrt{2/\alpha}$ is then added, where $\sigma_x=0.01$ and $\alpha=\Delta t/\tau$.

The output has 33 components: one fixation unit and a 32-unit direction ring. For each task, we generate 30 training batches of 1024 trials using a fixed random seed. Test trials are generated with a different seed and therefore do not overlap with the training trials.

\paragraph{Model architecture.} The model is a leaky-integrator continuous-time RNN with $N_{\mathrm h}=256$ hidden units. Its dynamics is described by the stochastic differential equation
\begin{equation}
  \tau\frac{d\mathbf h}{dt}
  =-\mathbf h
  +f\!\left(
    W^{\mathrm{in}}\mathbf x
    +W^{\mathrm{rec}}\mathbf h
    +\mathbf b
    +\sqrt{2\tau}\,\sigma_{\mathrm{rec}}\boldsymbol\xi(t)
  \right),
\end{equation}
where $\tau=100\,\mathrm{ms}$, $f$ is the softplus function, and each component of $\boldsymbol\xi(t)$ is a unit Gaussian white noise, with $\langle\xi_a(t)\xi_b(t')\rangle=\delta_{ab}\delta(t-t')$. The recurrent-noise scale is $\sigma_{\mathrm{rec}}=0.05$. With $\Delta t=20\,\mathrm{ms}$, $\alpha=\Delta t/\tau=0.2$. In numerical simulation, the equation is discretized as
\begin{equation}
  \mathbf h_{s+1}
  =(1-\alpha)\mathbf h_s
  +\alpha f\!\left(
    W^{\mathrm{in}}\mathbf x_s
    +W^{\mathrm{rec}}\mathbf h_s
    +\mathbf b
    +\sqrt{\frac{2}{\alpha}}\,
      \sigma_{\mathrm{rec}}\boldsymbol\xi_s
  \right),
  \qquad
  \boldsymbol\xi_s\sim\mathcal N(\mathbf 0,I).
\end{equation}
$W^{\mathrm{rec}}$ is initialized as a scaled identity matrix. At evaluation, we save the full hidden-state trajectory $(\mathbf h_1,\ldots,\mathbf h_S)$ for the drift analysis.

\paragraph{Loss function.} Let $\hat{\mathbf y}_{b,s}$ denote the network output and $\mathbf y_{b,s}$ the target for trial $b$ at time step $s$. The network minimizes the masked mean-squared error
\begin{equation}
  \mathcal L_{\mathrm{task}}
  =\left\langle
    m_{i,s}\bigl(\hat y_{b,s,i}-y_{b,s,i}\bigr)^2
  \right\rangle_{b,s,i},
\end{equation}
where $i$ indexes output units. The mask $m_{i,s}$ has weight 1 during fixation and stimulus presentation, 0 during the first 100\,ms of the response epoch, and 5 during the remainder of the response epoch. The weight of the fixation output is doubled to penalize premature responses.

\paragraph{Continual learning methods.} We implemented four continual-learning methods for the RNN. Naive sequential training, EWC, and experience replay follow the same principles as in the CNN experiments but use the trial-based mean-squared-error loss. The replay buffer stores 300 trials from each previous task. RNN-LwF uses the pre-task model outputs on the new-task trials as fixed targets and replaces the KL divergence with a mean-squared distillation loss. EWC and LwF did not achieve sufficiently reliable continual-learning performance for an interpretable drift comparison; the main RNN analysis therefore only reports naive training and replay.

\paragraph{Training.} Each task is trained with Adam \citep{kingma2015Adam} for at most 5,000 gradient steps, using a learning rate of $10^{-3}$ and a batch size of 1024. Gradients are computed by backpropagation through time over the full trial length $S$. Training stops early if the loss improvement remains below $10^{-3}$ for 500 consecutive steps. We save a model checkpoint and the probe-set hidden-state trajectories after every task.

\subsection{Notation and geometric measures}
\label{sec:notation}

Table~\ref{tab:notation} summarizes the main notations. The index $t=1,\ldots,T$ denotes both a task and the checkpoint saved after that task. Thus, $\theta_t$ denotes the model parameters after task $t$. The CNN has $T=20$ checkpoints and the RNN has $T=18$. The probe set $\mathcal P$ contains $P$ held-out samples from the first task and is fixed across checkpoints. Drift is measured by comparing the internal representations of these same samples at different checkpoints.

\begin{table}[ht]
\centering
\caption{Summary of notation.}
\label{tab:notation}
\small
\begin{tabular}{ll}
\toprule
Symbol & Meaning \\
\midrule
$T$ & Number of tasks and checkpoints (CNN: 20; RNN: 18) \\
$t$ & Task index, $t = 1,\ldots,T$ \\
$\theta_t$ & Model weights after task $t$ \\
$\mathcal{P}$ & Fixed probe set of $P$ samples \\
$P$ & Probe-set size \\
$D$ & CNN stage activation dimensionality \\
$N_{\mathrm{h}}$ & Number of RNN hidden units (= 256) \\
$S$ & Number of aligned time steps in an analyzed RNN batch \\
$B$ & Number of trials per probe batch \\
$b,\,s,\,n$ & Trial, time-step, and unit indices \\
$\tau$ & RNN membrane time constant (= 100\,ms) \\
\bottomrule
\end{tabular}
\end{table}

\paragraph{CNN representations.} For the CNN, the representation of probe sample $x\in\mathcal P$ at a given stage and checkpoint $t$ is the flattened activation vector
\begin{equation}
  \mathbf{r}_t(x) \;\in\; \mathbb{R}^{D},
\end{equation}
where $D$ is the number of elements in that stage's output. The drift vector relative to the first checkpoint is
\begin{equation}
  \boldsymbol{\delta}_t(x) = \mathbf{r}_t(x) - \mathbf{r}_1(x).
\end{equation}
The mean activation across the probe set is $\bar{\mathbf r}_t=P^{-1}\sum_{x\in\mathcal P}\mathbf r_t(x)$.

\paragraph{Representations in RNN} For an analyzed RNN probe batch with $B$ trials and $S$ aligned time steps, the hidden states at checkpoint $t$ form a tensor
\begin{equation}
  \mathcal{X}_t \;\in\; \mathbb{R}^{B \times S \times N_{\mathrm{h}}},
\end{equation}
indexed by trial $b$, time step $s$, and unit $n$. Following Deitch et al.\ \citep{deitch2021Representational}, we define four representations from this tensor. The population vector (PV) contains the activity of all units at one time step of one trial:
\begin{equation}
  \mathrm{PV}_{b,s}^{(t)}=\mathcal X_t[b,s,:]\in\mathbb R^{N_{\mathrm h}}.
\end{equation}
The ensemble rate vector (ERV) is the time-averaged activity of all units in one trial:
\begin{equation}
  \mathrm{ERV}_b^{(t)}
  =\frac{1}{S}\sum_{s=1}^{S}\mathcal X_t[b,s,:]
  \in\mathbb R^{N_{\mathrm h}}.
\end{equation}
The tuning curve vector (TCV) is the activity of one unit over a trial:
\begin{equation}
  \mathrm{TCV}_{b,n}^{(t)}=\mathcal X_t[b,:,n]\in\mathbb R^S.
\end{equation}
Finally, the spatiotemporal population vector (STPV) concatenates the full hidden-state trajectory of one trial:
\begin{equation}
  \mathrm{STPV}_b^{(t)}
  =\operatorname{vec}\!\bigl(\mathcal X_t[b,:,:]\bigr)
  \in\mathbb R^{S N_{\mathrm h}}.
\end{equation}
The STPV is the primary RNN representation analogous to that in the CNN analyses. Its drift vector relative to the first checkpoint is
\begin{equation}
  \boldsymbol\delta_t^{\mathrm{STPV}}(b)
  =\mathrm{STPV}_b^{(t)}-\mathrm{STPV}_b^{(1)}.
\end{equation}

\paragraph{Representational similarity measures and matrices.} For CNN stages, we use two complementary similarity summaries. First, to measure how the representation of the same probe image changes between checkpoints $i$ and $j$, we compare the matching activation vectors with cosine similarity and Euclidean distance:
\begin{equation}
  \operatorname{cos}(x;\,i,j) = \frac{\mathbf{r}_i(x)\cdot \mathbf{r}_j(x)}
    {\lVert \mathbf{r}_i(x)\rVert_2\,\lVert \mathbf{r}_j(x)\rVert_2}, \qquad
  \mathrm{dist}(x;\,i,j) = \lVert \mathbf{r}_i(x) - \mathbf{r}_j(x)\rVert_2.
\end{equation}
We average each measure over the fixed probe set and report its standard deviation across probe samples. Applying the cosine measure to all checkpoint pairs gives the model pairwise similarity matrix $M \in \mathbb{R}^{T\times T}$,
\begin{equation}
  M_{i,j} = \frac{1}{P}\sum_{x\in\mathcal{P}} \operatorname{cos}(x;\,i,j).
\end{equation}
This matrix summarizes how similar the representation at checkpoint $i$ is to the representation at checkpoint $j$ on the same probe images; grouping entries by task gap $|j-i|$ gives the drift-versus-gap curve. In particular, we write
$\langle\cos\rangle_t = P^{-1}\sum_{x\in\mathcal{P}}\operatorname{cos}(x;1,t)$
for the mean cosine similarity of the probe set relative to the first checkpoint, and
report the final drift $1-\langle\cos\rangle_T$ (averaged over the stages used in the
analysis) as a scalar summary of cumulative representational change.

Second, to describe the geometry among probe images within a single checkpoint, we compute the sample-wise Gram matrix $G^{(t)} \in \mathbb{R}^{P\times P}$,
\begin{equation}
  G_{ab}^{(t)} = \operatorname{cos}\!\bigl(\mathbf{r}_t(x_a),\,\mathbf{r}_t(x_b)\bigr).
\end{equation}
Unlike $M$, which compares the same image across checkpoints, $G^{(t)}$ compares different probe images within checkpoint $t$.
To assist visualization, probe samples are ordered by class so that within-class similarities form blocks along the diagonal in the similarity matrices.

\paragraph{Population vector correlations.} For each RNN representation type, we calculate the Pearson correlation between matching vectors at two checkpoints. For equal-length vectors $\mathbf u$ and $\mathbf v$,
\begin{equation}
  \rho(\mathbf{u}, \mathbf{v}) =
    \frac{(\mathbf{u} - \bar{u})\cdot(\mathbf{v} - \bar{v})}
         {\lVert \mathbf{u} - \bar{u}\rVert_2\,\lVert \mathbf{v} - \bar{v}\rVert_2},
\end{equation}
where $\bar u$ and $\bar v$ are the means of the vector elements. We then average over trials and, where applicable, time steps or units:
\begin{align}
  \rho_{\mathrm{PV}}(i, j) &= \frac{1}{B\, S}
    \sum_{b=1}^{B}\sum_{s=1}^{S}
    \rho\!\bigl(\mathrm{PV}_{b,s}^{(i)},\mathrm{PV}_{b,s}^{(j)}\bigr), \\
  \rho_{\mathrm{ERV}}(i, j) &= \frac{1}{B}
    \sum_{b=1}^{B}\rho\!\bigl(\mathrm{ERV}_b^{(i)},\mathrm{ERV}_b^{(j)}\bigr), \\
  \rho_{\mathrm{TCV}}(i, j) &= \frac{1}{B\, N_{\mathrm{h}}}
    \sum_{b=1}^{B}\sum_{n=1}^{N_{\mathrm h}}
    \rho\!\bigl(\mathrm{TCV}_{b,n}^{(i)},\mathrm{TCV}_{b,n}^{(j)}\bigr), \\
  \rho_{\mathrm{STPV}}(i, j) &= \frac{1}{B}
    \sum_{b=1}^{B}\rho\!\bigl(\mathrm{STPV}_b^{(i)},\mathrm{STPV}_b^{(j)}\bigr).
\end{align}
For representation type $v\in\{\mathrm{PV},\mathrm{ERV},\mathrm{TCV},\mathrm{STPV}\}$, the mean correlation at task gap $d$ is
\begin{equation}
  \bar\rho_v(d)=\frac{1}{T-d}\sum_{i=1}^{T-d}\rho_v(i,i+d),
  \qquad d=0,\ldots,T-1.
\end{equation}
For the CNN, the PV is the activation vector $\mathbf r_t(x)$. Thus, $\rho_{\mathrm{PV}}(i,j)$ is the mean correlation between matching sample activations, and $\rho_{\mathrm{ERV}}(i,j)$ is the correlation between the probe-set means $\bar{\mathbf r}_i$ and $\bar{\mathbf r}_j$.

\paragraph{UMAP visualization.} We use uniform manifold approximation and projection (UMAP) \citep{mcinnes2018UMAP} only for qualitative visualization; quantitative claims rely on the metrics defined above. For the CNN, we concatenate the probe representations from all checkpoints and apply principal component analysis (PCA). We retain 512 components (explaining approximately 70--90\% of the variance, depending on the continual-learning method), then fit one two-dimensional UMAP embedding to the PCA scores. The same fitted PCA and UMAP transformations are therefore used for every checkpoint. We split the resulting coordinates by checkpoint for plotting.

\paragraph{Replicates and uncertainty.} We trained 10 independently initialized networks for each reported condition. Drift and performance summaries were first computed within each network and then averaged across networks. Figure captions identify the uncertainty interval shown for each analysis.

\paragraph{Data and code availability.} The ImageNet-1k dataset is publicly available \citep{deng2009ImageNet,russakovsky2015ImageNet}. The cognitive-task trials are generated procedurally following \citet{yang2019Task}, as described above. All code for model training, analysis, and figure generation is available at \url{https://github.com/yksi2023/cl-drift}.

\printbibliography[title={References}]

@article{aitken2022Geometry,
  title = {The Geometry of Representational Drift in Natural and Artificial Neural Networks},
  author = {Aitken, Kyle and Garrett, Marina and Olsen, Shawn and Mihalas, Stefan},
  date = {2022-11-28},
  journaltitle = {PLOS Computational Biology},
  shortjournal = {PLOS Comput. Biol.},
  volume = {18},
  number = {11},
  pages = {e1010716},
  doi = {10.1371/journal.pcbi.1010716},
  url = {https://journals.plos.org/ploscompbiol/article?id=10.1371/journal.pcbi.1010716}
}

@online{chaudhry2019Tiny,
  title = {On Tiny Episodic Memories in Continual Learning},
  author = {Chaudhry, Arslan and Rohrbach, Marcus and Elhoseiny, Mohamed and Ajanthan, Thalaiyasingam and Dokania, Puneet K. and Torr, Philip H. S. and Ranzato, Marc'Aurelio},
  date = {2019-06-04},
  eprint = {1902.10486},
  eprinttype = {arXiv},
  doi = {10.48550/arXiv.1902.10486},
  url = {http://arxiv.org/abs/1902.10486}
}

@article{climer2025Hippocampal,
  title = {Hippocampal Representations Drift in Stable Multisensory Environments},
  author = {Climer, Jason R. and Davoudi, Heydar and Oh, Jun Young and Dombeck, Daniel A.},
  date = {2025-09-11},
  journaltitle = {Nature},
  shortjournal = {Nature},
  volume = {645},
  number = {8080},
  pages = {457--465},
  doi = {10.1038/s41586-025-09245-y},
  url = {https://www.nature.com/articles/s41586-025-09245-y}
}

@article{deitch2021Representational,
  title = {Representational Drift in the Mouse Visual Cortex},
  author = {Deitch, Daniel and Rubin, Alon and Ziv, Yaniv},
  date = {2021-10-11},
  journaltitle = {Current Biology},
  shortjournal = {Curr. Biol.},
  volume = {31},
  number = {19},
  pages = {4327-4339.e6},
  doi = {10.1016/j.cub.2021.07.062},
  url = {https://www.cell.com/current-biology/abstract/S0960-9822(21)01052-6}
}

@inproceedings{deng2009ImageNet,
  title = {{{ImageNet}}: A Large-Scale Hierarchical Image Database},
  shorttitle = {{{ImageNet}}},
  booktitle = {2009 {{IEEE Conference}} on {{Computer Vision}} and {{Pattern Recognition}}},
  author = {Deng, Jia and Dong, Wei and Socher, Richard and Li, Li-Jia and Li, Kai and Fei-Fei, Li},
  date = {2009-06},
  pages = {248--255},
  doi = {10.1109/CVPR.2009.5206848},
  url = {https://ieeexplore.ieee.org/document/5206848},
  eventtitle = {2009 {{IEEE Conference}} on {{Computer Vision}} and {{Pattern Recognition}}}
}

@article{dohare2024Loss,
  title = {Loss of Plasticity in Deep Continual Learning},
  author = {Dohare, Shibhansh and Hernandez-Garcia, J. Fernando and Lan, Qingfeng and Rahman, Parash and Mahmood, A. Rupam and Sutton, Richard S.},
  date = {2024-08},
  journaltitle = {Nature},
  shortjournal = {Nature},
  volume = {632},
  number = {8026},
  pages = {768--774},
  doi = {10.1038/s41586-024-07711-7},
  url = {https://www.nature.com/articles/s41586-024-07711-7}
}

@article{driscoll2017Dynamic,
  title = {Dynamic Reorganization of Neuronal Activity Patterns in Parietal Cortex},
  author = {Driscoll, Laura N. and Pettit, Noah L. and Minderer, Matthias and Chettih, Selmaan N. and Harvey, Christopher D.},
  date = {2017-08-24},
  journaltitle = {Cell},
  shortjournal = {Cell},
  volume = {170},
  number = {5},
  pages = {986-999.e16},
  doi = {10.1016/j.cell.2017.07.021},
  url = {https://www.cell.com/cell/abstract/S0092-8674(17)30828-0}
}

@article{driscoll2022Representational,
  title = {Representational Drift: Emerging Theories for Continual Learning and Experimental Future Directions},
  shorttitle = {Representational Drift},
  author = {Driscoll, Laura N. and Duncker, Lea and Harvey, Christopher D.},
  date = {2022-10-01},
  journaltitle = {Current Opinion in Neurobiology},
  shortjournal = {Curr. Opin. Neurobiol.},
  volume = {76},
  pages = {102609},
  doi = {10.1016/j.conb.2022.102609},
  url = {https://www.sciencedirect.com/science/article/pii/S0959438822001039}
}

@inproceedings{duncker2020Organizing,
  title = {Organizing Recurrent Network Dynamics by Task-Computation to Enable Continual Learning},
  booktitle = {Advances in {{Neural Information Processing Systems}}},
  author = {Duncker, Lea and Driscoll, Laura and Shenoy, Krishna V and Sahani, Maneesh and Sussillo, David},
  date = {2020},
  volume = {33},
  pages = {14387--14397},
  url = {https://proceedings.neurips.cc/paper/2020/hash/a576eafbce762079f7d1f77fca1c5cc2-Abstract.html}
}

@article{eppler2026Representational,
  title = {Representational Drift Reflects Ongoing Balancing of Stochastic Changes by Hebbian Learning},
  author = {Eppler, Jens-Bastian and Lai, Thomas and Aschauer, Dominik F. and Rumpel, Simon and Kaschube, Matthias},
  date = {2026-02-03},
  journaltitle = {Proceedings of the National Academy of Sciences},
  shortjournal = {Proc. Natl. Acad. Sci.},
  volume = {123},
  number = {5},
  pages = {e2503046123},
  doi = {10.1073/pnas.2503046123},
  url = {https://www.pnas.org/doi/10.1073/pnas.2503046123}
}

@article{gallego2020Longterm,
  title = {Long-Term Stability of Cortical Population Dynamics Underlying Consistent Behavior},
  author = {Gallego, Juan A. and Perich, Matthew G. and Chowdhury, Raeed H. and Solla, Sara A. and Miller, Lee E.},
  date = {2020-02},
  journaltitle = {Nature Neuroscience},
  shortjournal = {Nat. Neurosci.},
  volume = {23},
  number = {2},
  pages = {261--270},
  doi = {10.1038/s41593-019-0555-4},
  url = {https://www.nature.com/articles/s41593-019-0555-4}
}

@article{geva2023Time,
  title = {Time and Experience Differentially Affect Distinct Aspects of Hippocampal Representational Drift},
  author = {Geva, Nitzan and Deitch, Daniel and Rubin, Alon and Ziv, Yaniv},
  date = {2023-08-02},
  journaltitle = {Neuron},
  shortjournal = {Neuron},
  volume = {111},
  number = {15},
  pages = {2357-2366.e5},
  doi = {10.1016/j.neuron.2023.05.005},
  url = {https://www.cell.com/neuron/abstract/S0896-6273(23)00378-1}
}

@article{gonzalez2019Persistence,
  title = {Persistence of Neuronal Representations through Time and Damage in the Hippocampus},
  author = {Gonzalez, Walter G. and Zhang, Hanwen and Harutyunyan, Anna and Lois, Carlos},
  date = {2019-08-23},
  journaltitle = {Science},
  shortjournal = {Science},
  volume = {365},
  number = {6455},
  pages = {821--825},
  doi = {10.1126/science.aav9199},
  url = {https://www.science.org/doi/10.1126/science.aav9199}
}

@inproceedings{he2015Delving,
  title = {Delving Deep into Rectifiers: Surpassing Human-Level Performance on {{ImageNet}} Classification},
  shorttitle = {Delving Deep into Rectifiers},
  booktitle = {2015 {{IEEE International Conference}} on {{Computer Vision}} ({{ICCV}})},
  author = {He, Kaiming and Zhang, Xiangyu and Ren, Shaoqing and Sun, Jian},
  date = {2015-12},
  pages = {1026--1034},
  doi = {10.1109/ICCV.2015.123},
  url = {https://ieeexplore.ieee.org/document/7410480},
  eventtitle = {2015 {{IEEE International Conference}} on {{Computer Vision}} ({{ICCV}})}
}

@inproceedings{he2016Deep,
  title = {Deep {{Residual Learning}} for {{Image Recognition}}},
  booktitle = {Proceedings of the {{IEEE Conference}} on {{Computer Vision}} and {{Pattern Recognition}}},
  author = {He, Kaiming and Zhang, Xiangyu and Ren, Shaoqing and Sun, Jian},
  date = {2016},
  pages = {770--778},
  url = {https://www.cv-foundation.org/openaccess/content_cvpr_2016/html/He_Deep_Residual_Learning_CVPR_2016_paper.html?utm_source=chatgpt.com},
  eventtitle = {Proceedings of the {{IEEE Conference}} on {{Computer Vision}} and {{Pattern Recognition}}}
}

@online{hinton2015Distilling,
  title = {Distilling the {{Knowledge}} in a {{Neural Network}}},
  author = {Hinton, Geoffrey and Vinyals, Oriol and Dean, Jeff},
  date = {2015-03-09},
  eprint = {1503.02531},
  eprinttype = {arXiv},
  doi = {10.48550/arXiv.1503.02531},
  url = {http://arxiv.org/abs/1503.02531}
}

@article{keinath2022Representation,
  title = {The Representation of Context in Mouse Hippocampus Is Preserved despite Neural Drift},
  author = {Keinath, Alexandra T. and Mosser, Coralie-Anne and Brandon, Mark P.},
  date = {2022-05-03},
  journaltitle = {Nature Communications},
  shortjournal = {Nat. Commun.},
  volume = {13},
  number = {1},
  pages = {2415},
  doi = {10.1038/s41467-022-30198-7},
  url = {https://www.nature.com/articles/s41467-022-30198-7}
}

@inproceedings{kingma2015Adam,
  title = {Adam: {{A Method}} for {{Stochastic Optimization}}},
  shorttitle = {Adam},
  booktitle = {International {{Conference}} on {{Learning Representations}}},
  author = {Kingma, Diederik P. and Ba, Jimmy Lei},
  date = {2015},
  eprint = {1412.6980},
  eprinttype = {arXiv}
}

@article{kirkpatrick2017Overcoming,
  title = {Overcoming Catastrophic Forgetting in Neural Networks},
  author = {Kirkpatrick, James and Pascanu, Razvan and Rabinowitz, Neil and Veness, Joel and Desjardins, Guillaume and Rusu, Andrei A. and Milan, Kieran and Quan, John and Ramalho, Tiago and Grabska-Barwinska, Agnieszka and Hassabis, Demis and Clopath, Claudia and Kumaran, Dharshan and Hadsell, Raia},
  date = {2017-03-28},
  journaltitle = {Proceedings of the National Academy of Sciences},
  shortjournal = {Proc. Natl. Acad. Sci.},
  volume = {114},
  number = {13},
  pages = {3521--3526},
  doi = {10.1073/pnas.1611835114},
  url = {https://www.pnas.org/doi/10.1073/pnas.1611835114}
}

@inproceedings{kornblith2019Similarity,
  title = {Similarity of Neural Network Representations Revisited},
  booktitle = {Proceedings of the 36th {{International Conference}} on {{Machine Learning}}},
  author = {Kornblith, Simon and Norouzi, Mohammad and Lee, Honglak and Hinton, Geoffrey},
  date = {2019-05-24},
  pages = {3519--3529},
  url = {https://proceedings.mlr.press/v97/kornblith19a.html},
  eventtitle = {International {{Conference}} on {{Machine Learning}}}
}

@article{li2018Learning,
  title = {Learning without Forgetting},
  author = {Li, Zhizhong and Hoiem, Derek},
  date = {2018-12},
  journaltitle = {IEEE Transactions on Pattern Analysis and Machine Intelligence},
  shortjournal = {IEEE Trans. Pattern Anal. Mach. Intell.},
  volume = {40},
  number = {12},
  pages = {2935--2947},
  doi = {10.1109/TPAMI.2017.2773081},
  url = {https://ieeexplore.ieee.org/document/8107520}
}

@article{marks2021Stimulusdependent,
  title = {Stimulus-Dependent Representational Drift in Primary Visual Cortex},
  author = {Marks, Tyler D. and Goard, Michael J.},
  date = {2021-08-27},
  journaltitle = {Nature Communications},
  shortjournal = {Nat. Commun.},
  volume = {12},
  number = {1},
  pages = {5169},
  doi = {10.1038/s41467-021-25436-3},
  url = {https://www.nature.com/articles/s41467-021-25436-3}
}

@article{mcclelland1995Why,
  title = {Why There Are Complementary Learning Systems in the Hippocampus and Neocortex: Insights from the Successes and Failures of Connectionist Models of Learning and Memory},
  shorttitle = {Why There Are Complementary Learning Systems in the Hippocampus and Neocortex},
  author = {McClelland, James L. and McNaughton, Bruce L. and O'Reilly, Randall C.},
  date = {1995},
  journaltitle = {Psychological Review},
  shortjournal = {Psychol. Rev.},
  volume = {102},
  number = {3},
  pages = {419--457},
  doi = {10.1037/0033-295X.102.3.419}
}

@online{mcinnes2018UMAP,
  title = {{{UMAP}}: Uniform Manifold Approximation and Projection for Dimension Reduction},
  shorttitle = {Umap},
  author = {McInnes, Leland and Healy, John and Melville, James},
  date = {2018-02-09},
  eprint = {1802.03426},
  eprinttype = {arXiv},
  doi = {10.48550/arXiv.1802.03426},
  url = {http://arxiv.org/abs/1802.03426}
}

@article{micou2026Statistics,
  title = {Statistics of Cortical Representational Drift Can Enable Robust Readout},
  author = {Micou, Charles and O'Leary, Timothy},
  date = {2026-06-08},
  journaltitle = {PLOS Computational Biology},
  shortjournal = {PLOS Comput. Biol.},
  volume = {22},
  number = {6},
  pages = {e1014297},
  doi = {10.1371/journal.pcbi.1014297},
  url = {https://journals.plos.org/ploscompbiol/article?id=10.1371/journal.pcbi.1014297}
}

@article{mongillo2017Intrinsic,
  title = {Intrinsic Volatility of Synaptic Connections --- a Challenge to the Synaptic Trace Theory of Memory},
  author = {Mongillo, Gianluigi and Rumpel, Simon and Loewenstein, Yonatan},
  date = {2017-10-01},
  journaltitle = {Current Opinion in Neurobiology},
  shortjournal = {Curr. Opin. Neurobiol.},
  volume = {46},
  pages = {7--13},
  doi = {10.1016/j.conb.2017.06.006},
  url = {https://www.sciencedirect.com/science/article/pii/S0959438817300673}
}

@article{morales2025Representational,
  title = {Representational Drift and Learning-Induced Stabilization in the Piriform Cortex},
  author = {Morales, Guillermo B. and Muñoz, Miguel A. and Tu, Yuhai},
  date = {2025-07-22},
  journaltitle = {Proceedings of the National Academy of Sciences},
  shortjournal = {Proc. Natl. Acad. Sci.},
  volume = {122},
  number = {29},
  pages = {e2501811122},
  doi = {10.1073/pnas.2501811122},
  url = {https://www.pnas.org/doi/10.1073/pnas.2501811122}
}

@article{noda2025Homeostasis,
  title = {Homeostasis of a Representational Map in the Neocortex},
  author = {Noda, Takahiro and Kienle, Eike and Eppler, Jens-Bastian and Aschauer, Dominik F. and Kaschube, Matthias and Loewenstein, Yonatan and Rumpel, Simon},
  date = {2025-07},
  journaltitle = {Nature Neuroscience},
  shortjournal = {Nat. Neurosci.},
  volume = {28},
  number = {7},
  pages = {1533--1545},
  doi = {10.1038/s41593-025-01982-7},
  url = {https://www.nature.com/articles/s41593-025-01982-7}
}

@online{peters2026Coordinated,
  title = {Coordinated Representational Drift across the Mouse Cortex},
  author = {Peters, Ryan and Hope, James and Feldkamp, Michael and Beckerle, Travis and Oladepo, Ibrahim and Hryb, Ihor and Saxena, Kapil and Redish, A. David and Kodandaramaiah, Suhasa},
  date = {2026-05-09},
  eprint = {2026.05.05.723038},
  eprinttype = {bioRxiv},
  doi = {10.64898/2026.05.05.723038},
  url = {https://www.biorxiv.org/content/10.64898/2026.05.05.723038v1}
}

@online{qiao2019MicroBatch,
  title = {Micro-{{Batch Training}} with {{Batch-Channel Normalization}} and {{Weight Standardization}}},
  author = {Qiao, Siyuan and Wang, Huiyu and Liu, Chenxi and Shen, Wei and Yuille, Alan},
  date = {2019-03-25},
  eprint = {1903.10520},
  eprinttype = {arXiv},
  doi = {10.48550/arXiv.1903.10520},
  url = {http://arxiv.org/abs/1903.10520}
}

@article{qin2023Coordinated,
  title = {Coordinated Drift of Receptive Fields in Hebbian/Anti-Hebbian Network Models during Noisy Representation Learning},
  author = {Qin, Shanshan and Farashahi, Shiva and Lipshutz, David and Sengupta, Anirvan M. and Chklovskii, Dmitri B. and Pehlevan, Cengiz},
  date = {2023-02},
  journaltitle = {Nature Neuroscience},
  shortjournal = {Nat. Neurosci.},
  volume = {26},
  number = {2},
  pages = {339--349},
  doi = {10.1038/s41593-022-01225-z},
  url = {https://www.nature.com/articles/s41593-022-01225-z}
}

@article{ratzon2024Representational,
  title = {Representational Drift as a Result of Implicit Regularization},
  author = {Ratzon, Aviv and Derdikman, Dori and Barak, Omri},
  date = {2024-05-02},
  journaltitle = {eLife},
  shortjournal = {eLife},
  volume = {12},
  pages = {RP90069},
  doi = {10.7554/eLife.90069},
  url = {https://doi.org/10.7554/eLife.90069}
}

@article{rule2019Causes,
  title = {Causes and Consequences of Representational Drift},
  author = {Rule, Michael E and O'Leary, Timothy and Harvey, Christopher D},
  date = {2019-10-01},
  journaltitle = {Current Opinion in Neurobiology},
  shortjournal = {Curr. Opin. Neurobiol.},
  volume = {58},
  pages = {141--147},
  doi = {10.1016/j.conb.2019.08.005},
  url = {https://www.sciencedirect.com/science/article/pii/S0959438819300303}
}

@article{rule2022Selfhealing,
  title = {Self-Healing Codes: How Stable Neural Populations Can Track Continually Reconfiguring Neural Representations},
  shorttitle = {Self-Healing Codes},
  author = {Rule, Michael E. and O'Leary, Timothy},
  date = {2022-02-15},
  journaltitle = {Proceedings of the National Academy of Sciences},
  shortjournal = {Proc. Natl. Acad. Sci.},
  volume = {119},
  number = {7},
  pages = {e2106692119},
  doi = {10.1073/pnas.2106692119},
  url = {https://www.pnas.org/doi/10.1073/pnas.2106692119}
}

@article{russakovsky2015ImageNet,
  title = {{{ImageNet}} Large Scale Visual Recognition Challenge},
  author = {Russakovsky, Olga and Deng, Jia and Su, Hao and Krause, Jonathan and Satheesh, Sanjeev and Ma, Sean and Huang, Zhiheng and Karpathy, Andrej and Khosla, Aditya and Bernstein, Michael and Berg, Alexander C. and Fei-Fei, Li},
  date = {2015-12-01},
  journaltitle = {International Journal of Computer Vision},
  shortjournal = {Int. J. Comput. Vision},
  volume = {115},
  number = {3},
  pages = {211--252},
  doi = {10.1007/s11263-015-0816-y},
  url = {https://doi.org/10.1007/s11263-015-0816-y}
}

@article{schoonover2021Representational,
  title = {Representational Drift in Primary Olfactory Cortex},
  author = {Schoonover, Carl E. and Ohashi, Sarah N. and Axel, Richard and Fink, Andrew J. P.},
  date = {2021-06},
  journaltitle = {Nature},
  shortjournal = {Nature},
  volume = {594},
  number = {7864},
  pages = {541--546},
  doi = {10.1038/s41586-021-03628-7},
  url = {https://www.nature.com/articles/s41586-021-03628-7}
}

@online{sylte2025Coordinated,
  title = {Coordinated Representational Drift Supports Stable Place Coding in Hippocampal {{CA1}}},
  author = {Sylte, Ole Christian and Kilias, Antje and Bartos, Marlene and Sauer, Jonas-Frederic},
  date = {2025-02-05},
  eprint = {2025.02.04.636428},
  eprinttype = {bioRxiv},
  doi = {10.1101/2025.02.04.636428},
  url = {https://www.biorxiv.org/content/10.1101/2025.02.04.636428v1}
}

@article{vanderveldt2026Learning,
  title = {Learning Continually with Representational Drift},
  author = {family=Veldt, given=Suzanne, prefix=van der, useprefix=true and family=Ven, given=Gido M., prefix=van de, useprefix=true and Moorman, Sanne and Etter, Guillaume},
  date = {2026-06-15},
  journaltitle = {Trends Open},
  shortjournal = {Trends OPEN},
  doi = {10.1016/j.treopn.2026.04.013},
  url = {https://www.sciencedirect.com/science/article/pii/S3117347026000222}
}

@article{vandeven2020Braininspired,
  title = {Brain-Inspired Replay for Continual Learning with Artificial Neural Networks},
  author = {family=Ven, given=Gido M., prefix=van de, useprefix=true and Siegelmann, Hava T. and Tolias, Andreas S.},
  date = {2020-08-13},
  journaltitle = {Nature Communications},
  shortjournal = {Nat. Commun.},
  volume = {11},
  number = {1},
  pages = {4069},
  doi = {10.1038/s41467-020-17866-2},
  url = {https://www.nature.com/articles/s41467-020-17866-2}
}

@article{vandeven2022Three,
  title = {Three Types of Incremental Learning},
  author = {family=Ven, given=Gido M., prefix=van de, useprefix=true and Tuytelaars, Tinne and Tolias, Andreas S.},
  date = {2022-12},
  journaltitle = {Nature Machine Intelligence},
  shortjournal = {Nat. Mach. Intell.},
  volume = {4},
  number = {12},
  pages = {1185--1197},
  doi = {10.1038/s42256-022-00568-3},
  url = {https://www.nature.com/articles/s42256-022-00568-3}
}

@inproceedings{wu2018Group,
  title = {Group {{Normalization}}},
  booktitle = {Proceedings of the {{European Conference}} on {{Computer Vision}}},
  author = {Wu, Yuxin and He, Kaiming},
  date = {2018},
  pages = {3--19},
  doi = {10.1007/978-3-030-01261-8_1},
  eventtitle = {Proceedings of the {{European Conference}} on {{Computer Vision ECCV}}},
  isbn = {978-3-030-01261-8}
}

@article{yang2019Task,
  title = {Task Representations in Neural Networks Trained to Perform Many Cognitive Tasks},
  author = {Yang, Guangyu Robert and Joglekar, Madhura R. and Song, H. Francis and Newsome, William T. and Wang, Xiao-Jing},
  date = {2019-02},
  journaltitle = {Nature Neuroscience},
  shortjournal = {Nat. Neurosci.},
  volume = {22},
  number = {2},
  pages = {297--306},
  doi = {10.1038/s41593-018-0310-2},
  url = {https://www.nature.com/articles/s41593-018-0310-2}
}

@inproceedings{zenke2017Continual,
  title = {Continual Learning through Synaptic Intelligence},
  booktitle = {Proceedings of the 34th {{International Conference}} on {{Machine Learning}}},
  author = {Zenke, Friedemann and Poole, Ben and Ganguli, Surya},
  date = {2017-07-17},
  pages = {3987--3995},
  url = {https://proceedings.mlr.press/v70/zenke17a.html},
  eventtitle = {International {{Conference}} on {{Machine Learning}}}
}

@article{ziv2013Longterm,
  title = {Long-Term Dynamics of {{CA1}} Hippocampal Place Codes},
  author = {Ziv, Yaniv and Burns, Laurie D. and Cocker, Eric D. and Hamel, Elizabeth O. and Ghosh, Kunal K. and Kitch, Lacey J. and Gamal, Abbas El and Schnitzer, Mark J.},
  date = {2013-03},
  journaltitle = {Nature Neuroscience},
  shortjournal = {Nat Neurosci},
  volume = {16},
  number = {3},
  pages = {264--266},
  doi = {10.1038/nn.3329},
  url = {https://www.nature.com/articles/nn.3329}
}

\section{Appendix}

\renewcommand{\thefigure}{S\arabic{figure}}
\setcounter{figure}{0}

\subsection{Stability of sample-similarity structure measured with centered kernel alignment}
\label{sec:cka_appendix}

\begin{figure}[ht]
  \centering
  \begin{overpic}[width=0.92\linewidth]{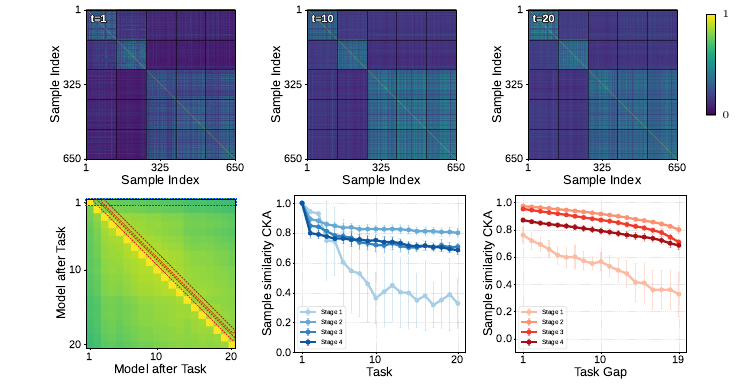}
    \put(5,52){{\textbf{\textsf{a}}}}
    \put(34,52){{\textbf{\textsf{b}}}}
    \put(65,52){{\textbf{\textsf{c}}}}
    \put(5,26){{\textbf{\textsf{d}}}}
    \put(34,26){{\textbf{\textsf{e}}}}
    \put(65,26){{\textbf{\textsf{f}}}}
  \end{overpic}
  \caption{Sample-similarity structure for the ImageNet CNN trained with replay.
  (a--c)~Pairwise cosine similarity of probe representations at residual stage~4 after tasks 1, 10, and 20, with samples ordered by class.
  (d)~Pairwise CKA of stage-4 sample-similarity matrices across all checkpoints.
  The blue dashed outline marks the first row (CKA relative to task~1), shown as a
  function of checkpoint in (e). The red dashed outlines mark the off-diagonals of
  gaps 1 and 2 as a schematic: each point in (f) is the mean CKA along one such
  off-diagonal of gap $d$.
  (e)~CKA relative to the task-1 similarity matrix versus checkpoint.
  (f)~Mean CKA as a function of task gap.
  Curves in (e) and (f) from light to dark show residual stages 1--4; error bars
  denote the standard deviation across 10 independently trained networks.}
  \label{fig:cka_similarity_gap}
\end{figure}

The drift measures in the main text compare the representation of the same probe sample at two checkpoints. Such first-order measures are sensitive to any coordinated transformation of the population code: a rotation of the representation displaces every activation vector even when the relations among samples are unchanged. To ask whether the geometry of the probe set is preserved while individual vectors drift, we examined the sample-wise similarity matrices $G^{(t)}$ defined in Methods.
Each entry $G_{ab}^{(t)}$ is the cosine similarity between the representations $\mathbf{r}_t(x_a)$ and $\mathbf{r}_t(x_b)$ of probe samples $x_a$ and $x_b$ at checkpoint $t$. For display, samples are ordered by class, so a stable class geometry appears as persistent blocks along the diagonal.
At residual stage~4 of the ImageNet network trained with replay, these matrices after tasks 1, 10, and 20 retain a similar block structure (Fig.~\ref{fig:cka_similarity_gap}a--c).

We then compared matrices across checkpoints with centered kernel alignment (CKA) \citep{kornblith2019Similarity}. Let $H=I_P-\frac{1}{P}\mathbf 1\mathbf 1^\top$ denote the centering matrix and $\tilde G^{(t)}=HG^{(t)}H$ the doubly centered similarity matrix. The alignment between checkpoints $i$ and $j$ is
\begin{equation}
  \mathrm{CKA}(i,j)=
  \frac{\bigl\langle \tilde G^{(i)},\,\tilde G^{(j)}\bigr\rangle_{\mathrm F}}
       {\bigl\lVert \tilde G^{(i)}\bigr\rVert_{\mathrm F}\,
        \bigl\lVert \tilde G^{(j)}\bigr\rVert_{\mathrm F}},
\end{equation}
where $\langle\cdot,\cdot\rangle_{\mathrm F}$ and $\lVert\cdot\rVert_{\mathrm F}$ denote the Frobenius inner product and norm. Because $G^{(t)}$ is the Gram matrix of the $\ell_2$-normalized activations, this quantity is the linear CKA of \citet{kornblith2019Similarity} applied to length-normalized representations. CKA equals 1 when two checkpoints induce identical similarity structure over the probe set and, unlike the first-order measures, is invariant to rotations, reflections, and isotropic rescaling of the representation.

Computing $\mathrm{CKA}(i,j)$ for every checkpoint pair at stage~4 yields the matrix in Fig.~\ref{fig:cka_similarity_gap}d. Two slices of this matrix are shown as curves. The blue dashed outline marks the first row---CKA of each later checkpoint with the task-1 matrix---which is plotted versus checkpoint in Fig.~\ref{fig:cka_similarity_gap}e. Each point in Fig.~\ref{fig:cka_similarity_gap}f is the mean CKA along one off-diagonal of gap $d=|j-i|$; the red dashed outlines mark only the $d=1$ and $d=2$ bands, as a schematic of this grouping rather than as the only gaps included. In residual stages 2--4, the similarity structure was markedly more stable than
the underlying activation vectors: CKA between adjacent checkpoints was 0.87--0.97
and declined only gradually with task separation, remaining at approximately
0.68--0.80 after 19 intervening tasks, whereas the sample-level population-vector
correlations fell substantially over the same gaps (Fig.~\ref{fig:cnn_method_grid}l).
Much of the drift is therefore consistent with transformations that approximately
preserve the pairwise geometry of the probe set, rather than with an unstructured
re-mixing of the representation. Among stages 2--4, deeper stages were less stable,
matching the depth dependence of the first-order measures. Stage 1 behaved
differently: its CKA was low even between adjacent checkpoints ($\approx 0.64$)
and highly variable across networks, indicating that the similarity structure of
the earliest stage is only weakly conserved across checkpoints and differs strongly
between independently trained networks.

These results provide quantitative support for the qualitative UMAP observation in the main text (Fig.~\ref{fig:cnn_umap}): under replay, the relational geometry among task-1 probe samples---the structure on which class readout relies---is largely preserved and changes only slowly with task separation, even as the individual probe representations drift.

\subsection{Temporal structure of RNN drift}

\begin{figure}[ht]
  \centering
  \begin{overpic}[width=0.8\linewidth]{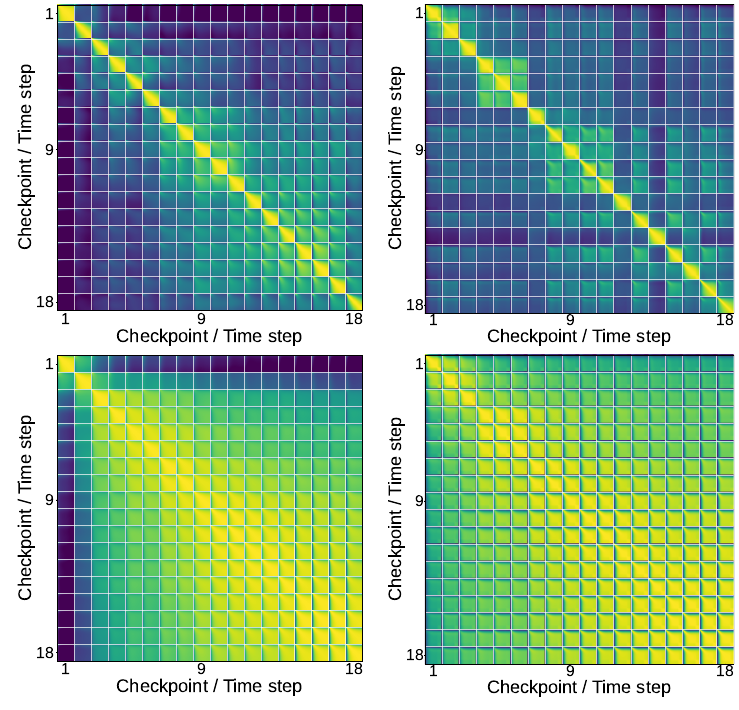}
    \put(1,96){{\small (a)}}
    \put(51,96){{\small (b)}}
    \put(1,47){{\small (c)}}
    \put(51,47){{\small (d)}}
  \end{overpic}
  \caption{Temporal structure of recurrent-network drift for the fdgo probe task. Rows
  show naive sequential training and replay; columns show fixation and the combined
  stimulus--response epochs. Panels: (a) naive training, fixation; (b) naive training,
  stimulus--response; (c) replay, fixation; and (d) replay, stimulus--response.}
  \label{fig:rnn_temporal_drift}
\end{figure}

We asked how recurrent-network drift was distributed over the course of a trial. For the probe task, we computed a cross-checkpoint, cross-time correlation matrix in which each entry is the Pearson correlation between the population vector at one checkpoint and time point and that at another (Fig.~\ref{fig:rnn_temporal_drift}). Diagonal blocks compare time points within one checkpoint and reveal the trial's temporal structure; off-diagonal blocks compare population states across checkpoints and reveal drift. We analyzed fixation (left column) separately from the stimulus and response epochs (right column).

Under naive sequential training, off-diagonal blocks were dark: states from distant checkpoints were nearly uncorrelated, reflecting catastrophic overwriting of the probe-task representation. Under replay, off-diagonal similarity remained higher, especially during the stimulus and response epochs. Preservation was graded rather than complete, however, and similarity declined with checkpoint separation. Meanwhile, the repeated block structure retained the trial's temporal organization: time points in corresponding epochs remained correlated, whereas unrelated epochs did not. Replay therefore allowed each epoch's population state to change gradually without dissolving the temporal structure of the computation.

\end{document}